\documentclass[submission, Proceedings]{SciPost}

\begin{document}

\begin{center}{\Large \textbf{
Non-standard interactions in  $\tau^- \to (\pi^-\eta,\pi^-\pi^0)\nu_{\tau}$ decays
}}\end{center}

\begin{center}
G. L\'opez Castro \textsuperscript{1}
\end{center}

\begin{center}
{\bf 1} Departamento de F\'\i sica, Centro de Investigaci\'on y de Estudios Avanzados, Apartado Postal 14-740, C.P. 07000 Ciudad de M\'exico, M\'exico
\\
email: glopez@fis.cinvestav.mx
\end{center}

\begin{center}
\today
\end{center}

\definecolor{palegray}{gray}{0.95}
\begin{center}
\colorbox{palegray}{
  \begin{tabular}{rr}
  \begin{minipage}{0.05\textwidth}
    \includegraphics[width=8mm]{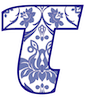}
  \end{minipage}
  &
  \begin{minipage}{0.82\textwidth}
    \begin{center}
    {\it Proceedings for the 15th International Workshop on Tau Lepton Physics,}\\
    {\it Amsterdam, The Netherlands, 24-28 September 2018} \\
    \href{https://scipost.org/SciPostPhysProc.1}{\small \sf scipost.org/SciPostPhysProc.Tau2018}\\
    \end{center}
  \end{minipage}
\end{tabular}
}
\end{center}


\section*{Abstract}
{\bf
Originally thought as clean processes to study the hadronization of the weak currents, semileptonic tau lepton decays can be useful to set constraints on non-standard (NS) weak interactions. In this talk we summarize our recent studies on the effects of NS interactions in $\tau^- \to (\pi^-\pi^0,\pi^-\eta)\nu_{\tau}$ decays. We find that experimental data on these decays provide strong  constraints on NS scalar and tensor interactions, respectively. Further improved measurements at Belle II and a better knowledge of necessary tensor and scalar form factors will allow to set limits on these NS interactions that are similar or better to contraints obtained from other low-energy processes.}

\vspace{10pt}
\noindent\rule{\textwidth}{1pt}
\tableofcontents\thispagestyle{fancy}
\noindent\rule{\textwidth}{1pt}
\vspace{10pt}

\section{Introduction}
\label{sec:intro}
Exclusive semileptonic decays of tau leptons provide a clean environment to study the hadronization of the charged weak currents up to the tau lepton mass scale \cite{Pich:2013lsa}. For the same reason, they are useful also to study the properties of light hadronic resonances and to extract information on the quark mixing parameters $V_{ud}$ and $V_{us}$. On the other hand, if we rely on the Standard Model (SM) description of the hadronization processes, semileptonic decays of tau leptons can be very useful in constraining the effects of non-standard (NS)  interactions.  Here  we summarize some  of our recent contributions  \cite{Garces:2017jpz, Miranda:2018cpf} which show that scalar and tensor interactions can be sensitively constrained from $\tau^- \to (\pi^-\pi^0,\pi^-\eta)\nu_{\tau}$ semileptonic decays, respectively. 

  The interest of $\tau^- \to P^-P^0\nu_{\tau}$ ($P$ a pseudoscalar meson) semileptonic decays in the search of effects of New Physics is not new. Forty years ago it was suggested \cite{Leroy:1977pq} that scalar (second class) currents \cite{Weinberg:1958ut}, if they exist, could manifest in $\tau^- \to \pi^-\eta\nu_{\tau}$ decays through the dominance of the $a_0(980)$ scalar meson. Despite strong experimental efforts, this process has not been observed so-far and the experimental upper limit of $9.9 \times 10^{-5}$ (95\% c.l.) \cite{delAmoSanchez:2010pc, Tanabashi:2018oca} has been set on its branching ratio. On the theoretical side, the different predictions for this decay spread over an order or magnitude (see Table 1 for recent predictions \cite{Nussinov:2008gx}-\cite{Escribano:2016ntp}), mainly because experimental information on the scalar form factor associated to the $\pi^-\eta$ system is not available. 
  
  \begin{table}[h!]
\centering
\begin{tabular}{|l|l|l|l|}
\hline
$B$(Vector)         & $B$(Scalar)  & $B$(sum) & Reference \\ \hline
$0.36$ & $1.0$ & $1.36$ & \cite{Nussinov:2008gx} \\ \hline
$0.33\sim 0.47$ & $1.73\sim 3.33$ & $2.1\sim 3.8$ & \cite{Paver:2010mz} \\ \hline
$0.44$ & $0.04$ & $0.48$ & \cite{Volkov:2012be} \\ \hline
$0.13$ & $0.20$ & $0.33$ & \cite{Descotes-Genon:2014tla} \\ \hline
$0.26$ & $1.41$ & $1.67$ & \cite{Escribano:2016ntp} \\ \hline
\end{tabular}
\caption{Predicted branching ratios (in $10^{-5}$ units) for $\tau^- \to \pi^-\eta\nu_{\tau}$ decays induced by isospin breaking $\pi^0-\eta$ mixing. Contributions of vector and scalar form factors to the branching ratios (and their sum) are shown separately. }\label{BRIn}
\end{table}
  
    Non-standard scalar and tensor interactions have been invoked as a mechanism to induce CP violation in $\tau \to K\pi\nu_{\tau}$ \cite{Delepine:2006fv}. Although these currents can indeed generate CP violation in semileptonic tau decays, tensor currents can not be responsible \cite{Cirigliano:2017tqn} for the large  CP rate asymmetry reported by BABAR \cite{BABAR:2011aa}.  

 This contribution presents a summary of our recent studies  on the effects of non-standard scalar and tensor interactions in $\tau^- \to \pi^-\eta\nu_{\tau}$ \cite{Garces:2017jpz} and $\tau^- \to \pi^-\pi^0\nu_{\tau}$ \cite{Miranda:2018cpf} decays. These two decays are very sensitive probes of NS interactions and provide complementary constraints on both types of NS couplings. A different approach to NS interactions in tau semileptonic decays has been presented in \cite{Cirigliano:2018dyk} with similar conclusions.

\section{NS interactions in semileptonic tau lepton decays}
  Effects of New Physics degrees of freedom in strangeness-conserving semileptonic $\tau$ lepton decays should manifest in the following dimension-6, lepton-flavor conserving effective Lagrangian \cite{Cirigliano:2009wk}:
  \begin{eqnarray}
 { \cal L}_{\rm eff}&= &-\sqrt{2}G_FV_{ud}N_{\epsilon} \left\{\bar{ \tau}_L\gamma_{\mu}\nu_L \cdot \bar{u}\gamma^{\mu}\left(1-[1-2\widehat{\epsilon}_R]\gamma_5 \right) d\frac{}{} \right. \nonumber \\
&& \left. + \frac{}{} \bar{\tau}_R \nu_L\cdot \bar{u}\left( \widehat{\epsilon}_S-\widehat{\epsilon}_P\gamma_5\right) d+2 \widehat{\epsilon}_T\bar{\tau}_R\sigma_{\mu\nu}\nu_L\cdot \bar{u}\sigma^{\mu\nu}d \right\} +{\rm h. c.},
  \end{eqnarray}
  where the subindices in the lepton fields refers to Left- or Right-handed chiralities and we have defined $N_{\epsilon}\equiv 1+\epsilon_L+\epsilon_R$, $\widehat{\epsilon}_i\equiv \epsilon_i/N_{\epsilon}$  ($i=R, S, P, T$) and $\sigma_{\mu\nu}=i[\gamma_{\mu},\gamma_{\nu}]/2$.  The SM is recovered when  the dimensionless Wilson coefficients $\epsilon_{L, R, S,P,T}=0$ vanish.
  
    By including the effects of short-distance electroweak corrections $S_{EW}$ (its numerical value is not relevant here as it cancels in our observables), the decay amplitude for semileptonic $\tau^- \to \pi^-(p_-)P^0(p_0)\nu_{\tau}$ (with $P^0=\pi^0, \eta, \eta'$) reads:
    \begin{equation} \label{amp}
    {\cal M}=\frac{G_FV_{ud}\sqrt{S_{EW}}}{\sqrt{2}}N_{\epsilon}\left\{L_{\mu}\cdot  H^{\mu}+\widehat{\epsilon}_S L\cdot H+ 2\widehat{\epsilon}_T L_{\mu\nu}\cdot H^{\mu\nu}\right\}\ ,
    \end{equation}
where we have defined the leptonic currents $L_{\mu}\equiv \bar{u}_{\nu_{\tau}}\gamma_{\mu}(1-\gamma_5)u_{\tau}, \ L\equiv \bar{u}_{\nu_{\tau}}(1-\gamma_5)u_{\tau}, \L_{\mu\nu}\equiv \bar{u}_{\nu_{\tau}}\sigma_{\mu\nu}(1-\gamma_5)u_{\tau}$ and the hadronic matrix elements:
\begin{eqnarray} \label{hme}
H^{\mu} &=& \langle P^0\pi^- |\bar{d}\gamma^{\mu}u|0\rangle =F_+^{P^0\pi^-}(s) \left[ (p_0-p_-)^{\mu}-\frac{\Delta_{P^0\pi^-}}{s}q^{\mu}\right] +F_0^{P^0\pi^-}(s)\frac{\Delta_{P^0\pi^-}}{s}q^{\mu}, \nonumber \\
H &=&  \langle P^0\pi^- |\bar{d}u|0\rangle = S^{P^0\pi^-}(s), \\
H^{\mu\nu}&=&  \langle P^0\pi^- |\bar{d}\sigma^{\mu\nu}u|0\rangle = iF_T^{P^0\pi^-}(s) \left( p_0^{\mu}p_-^{\nu}- p_0^{\nu}p_-^{\mu}\right) \, \nonumber 
\end{eqnarray}
where we have defined $\Delta_{P^0\pi^-}\equiv m_{P^0}^2-m_{\pi^-}^2$. 

The four form factors involved in hadronic matrix elements depend on the momentum transfer squared  $s\equiv q^2=(p_0+p_-)^2$, within the kinematical range $(m_{P^0}+m_{\pi^-})^2 \leq s \leq m_{\tau}^2$. Owing to the divergence of the vector current $i\partial_{\mu}(\bar{d}\gamma^{\mu}u)=(m_d-m_u)\bar{d}u$, we can relate the true  $S(s)$ and induced $F_0(s)$ scalar form factors, in such a way that the former can be absorbed into the later giving rise to the following effective scalar form factor \cite{Garces:2017jpz,Miranda:2018cpf} to be replaced in the induced form factor in the decay amplitude \ref{amp}
\begin{equation} \label{sff}
F_0(s) \rightarrow F_S(s)\equiv F_0(s)\left( 1+ \frac{s\widehat{\epsilon}_S}{m_{\tau}(m_d-m_u)} \right)\ .
\end{equation}
It is interesting to note that the effects of NS scalar interactions grow linearly with $s$. On the other hand, in the case of $\pi^-\eta(\eta')$ final states, where the induced scalar form factor $F_0(s)$ is suppressed by isospin breaking, the effects of new scalar interactions is of $O(0)$ in the isospin breaking parameter. This allows us to expect that $\tau^- \to  \pi^-\pi^0\nu_{\tau}$ decays will be sensitive to tensor interactions, while the isospin breaking violating $\tau^- \to  \pi^-\eta(\eta')\nu_{\tau}$ decays can be sensitive to scalar interactions. In this way, both decays play a complementary role in constraining the non-standard scalar and tensor interactions.

\section{Form factor models}
Following the redefinition in Eq. (\ref{sff}),  each of the tau lepton decays under consideration depend upon three form factors. While experimental information is available for the vector form factor  below the tau mass scale, this is not the case for the scalar and tensor form factors. For the later we have to rely on theoretical models subject to general basic constraints. For definiteness hereafter we will refer only to the $\pi^-\pi^0$ and $\pi^-\eta$ decay channels of tau leptons.
\subsection{Vector form factor}
The vector form factor for $\tau^- \to \pi^-\eta\nu_{\tau}$ decay can be written as $F^{\eta\pi^-}_+(s)=\epsilon_{\pi\eta}F_+^{\pi^0\pi^-}(s)$ \cite{Escribano:2016ntp}, where $\epsilon_{\pi\eta}=(9.8\pm 0.3)\times 10^{-3}$ corresponds to the $\pi-\eta$ mixing `angle' and $F_+^{\pi^0\pi^-}(s)$ corresponds to the weak pion vector form factor. For the later we use the expression derived \cite{Dumm:2013zh} using from unitarized  chiral perturbation theory matched to resonances where the free parameters are fixed from the pion form factor obtained from Belle data \cite{Fujikawa:2008ma}. 
\subsection{Scalar form factor}
As discussed in the previous section, information on the scalar form factor is relevant only for the description of  $\tau^- \to \pi^-\eta\nu_{\tau}$ decays. The spread of predictions in Table \ref{BRIn} stems mainly from the different models for this form factor. In Ref \cite{Garces:2017jpz} we have used a three-times subtracted dispersive and unitarized representation of the scalar form factor where the phase is fixed from the three coupled channel analysis that are permited in $\pi^-\eta$ rescattering \cite{Escribano:2016ntp}. In Ref. \cite{Miranda:2018cpf} we follow the analysis of Ref. \cite{Descotes-Genon:2014tla} for the subleading $\pi\pi$ scalar form factor in the elastic approach.
\subsection{Tensor form factor}
The tensor form factor $F_T^{P^0\pi^-}(s)$ describing the hadronization of the tensor current  is obtained  (at zero recoil)\cite{Garces:2017jpz} from the leading chiral Lagrangian ${\cal L}^{O(p^4)}=\Lambda_1\langle t^{\mu\nu}_+f_{+\mu\nu}\rangle-i\Lambda_2\langle t^{\mu\nu}_+u_{\mu}u_{\nu}\rangle$ (see definitions in \cite{Cata:2007ns}): $F_T^{\eta\pi^-}(0)=\epsilon_{\pi\eta}F_T^{\pi^0\pi^-}(0)=\epsilon_{\pi\eta}\sqrt{2}\Lambda_2/F^2$. Based on dimensional arguments, we have estimated $F_T^{\eta\pi^-}(0) \leq 9.4 \times 10^{-2}$ GeV$^{-1}$ \cite{Garces:2017jpz}, which is about a factor 5 larger than a calculation based on lattice QCD ($\approx 1.96(0.11)\times 10^{-2}$ GeV$^{-1}$) \cite{Baum:2011rm}. Since the tensor form factor has a negligible effect in $\tau^- \to \pi^-\eta\nu$ decays, its precise normalization and $s$-dependence is not very relevant.

  A more precise description of the tensor form factor is required for $\tau^- \to \pi^-\pi^0\nu_{\tau}$ decays. Here, a dispersive representation of the form factor \cite{Miranda:2018cpf}
  \begin{equation}
  F_T^{\pi^0\pi^-}(s)=F_T^{\pi^0\pi^-}(0)\exp \left\{\frac{s}{\pi}\int_{4m_{\pi}^2}^{\infty} ds' \frac{\delta_T(s')}{s'(s'-s-i\epsilon)} \right\}
  \end{equation}
  looks more appropriate. In the elastic approximation, the phase can be approximated by the spin-1 $\rho$-resonance in the region below 1 GeV \cite{Miranda:2018cpf,Cirigliano:2018dyk}. For the normalization of the tensor form factor we use $F_T^{\pi^0\pi^-}(0)=(2.0\pm 0.1)$ GeV$^{-1}$ which is estimated from the tensor charge of the pion form factor evaluated in the lattice calculation of Ref. \cite{Baum:2011rm}.

\section{Decay observables}
\label{sec:another}
  
  In Refs. \cite{Garces:2017jpz,Miranda:2018cpf} we have studied different observables (Dalit Plot, hadronic-mass and angular distributions, forward-backward asymmetries of the charged pion, decay rates) associated to the $\tau \to (\pi^-\pi^0, \pi^-\eta, \pi^-\eta')\nu_{\tau}$ decays.  Although some of them exhibit some sensitivity to the effects of NS interactions, the most sensitive turns out to be the properly normalized decay rates. Therefore, in this contribution we will focus only on the results obtained from the decay rates for the $\pi^-\pi^0$ and $\pi^-\eta$ channels. 
  
  In the rest frame of the decaying tau lepton,  the hadronic mass distribution for the generic semileptonic decay is given by:
    \begin{equation} \label{hadmass}
    \frac{d\Gamma}{ds}=\frac{G_F^2N_{\epsilon}^2S_{EW}m_{\tau}^3\left| V_{ud}F_+^{n\pi^-}(0)\right|^2} {192\pi^3\sqrt{s}} \left( 1-\frac{s}{m_{\tau}^2}\right)^2 \left|{\bf p}_{\pi^-}\right| \left[X^{P^0\pi^-}_{VA+S}+ \widehat{\epsilon}_TX^{P^0\pi^-}_T+\widehat{\epsilon}_T^2 X^{P^0\pi^-}_{T^2}\right]\ ,
    \end{equation}
  where ${\bf p}_{\pi^-}$ is the three-momentum of the $\pi^-$ in the rest frame of the $\pi^-\eta$ system and 
  \begin{eqnarray} \label{coeff-seps}
  X^{P^0\pi^-}_{VA+S} &=& 4\left[ \left|\widetilde{ F}_+^{P^0\pi^-}(s)\right|^2 \left( 1+2\frac{2s}{m_{\tau}^2}\right)\frac{\left|{\bf p}_{\pi^-}\right|^2}{s}+ \frac{3\Delta_{P^0\pi^-}^2}{4s^2} \left| \widetilde{F}^{P^0\pi^-}_S(s)\right|^2 \right] \nonumber \\
  X^{P^0\pi^-}_T &=& -\frac{24\sqrt{2}{\rm Re}\left[ \widetilde{F}_+^{P^0\pi^-}(s)(\widetilde{F}_T^{P^0\pi^-}(s))^*\right]}{m_{\tau}} \left| {\bf p}_{\pi^-}\right|^2 \nonumber \\
  X^{P^0\pi^-}_{T^2} &=& 16 \left| \widetilde{F}_T^{P^0\pi^-}(s)\right|^2 \left( 1+ \frac{s}{2m_{\tau}^2}\right)  \left| {\bf p}_{\pi^-}\right|^2 
  \end{eqnarray}
  with $\widetilde{F}_i ^{P^0\pi^-}(s)=F_i^{P^0\pi^-}(s)/F_+^{P^0\pi^-}(0)$ and $i=+, S, T$. 
  
\begin{figure*}[h!]
\begin{center}
    \includegraphics[scale=0.50]{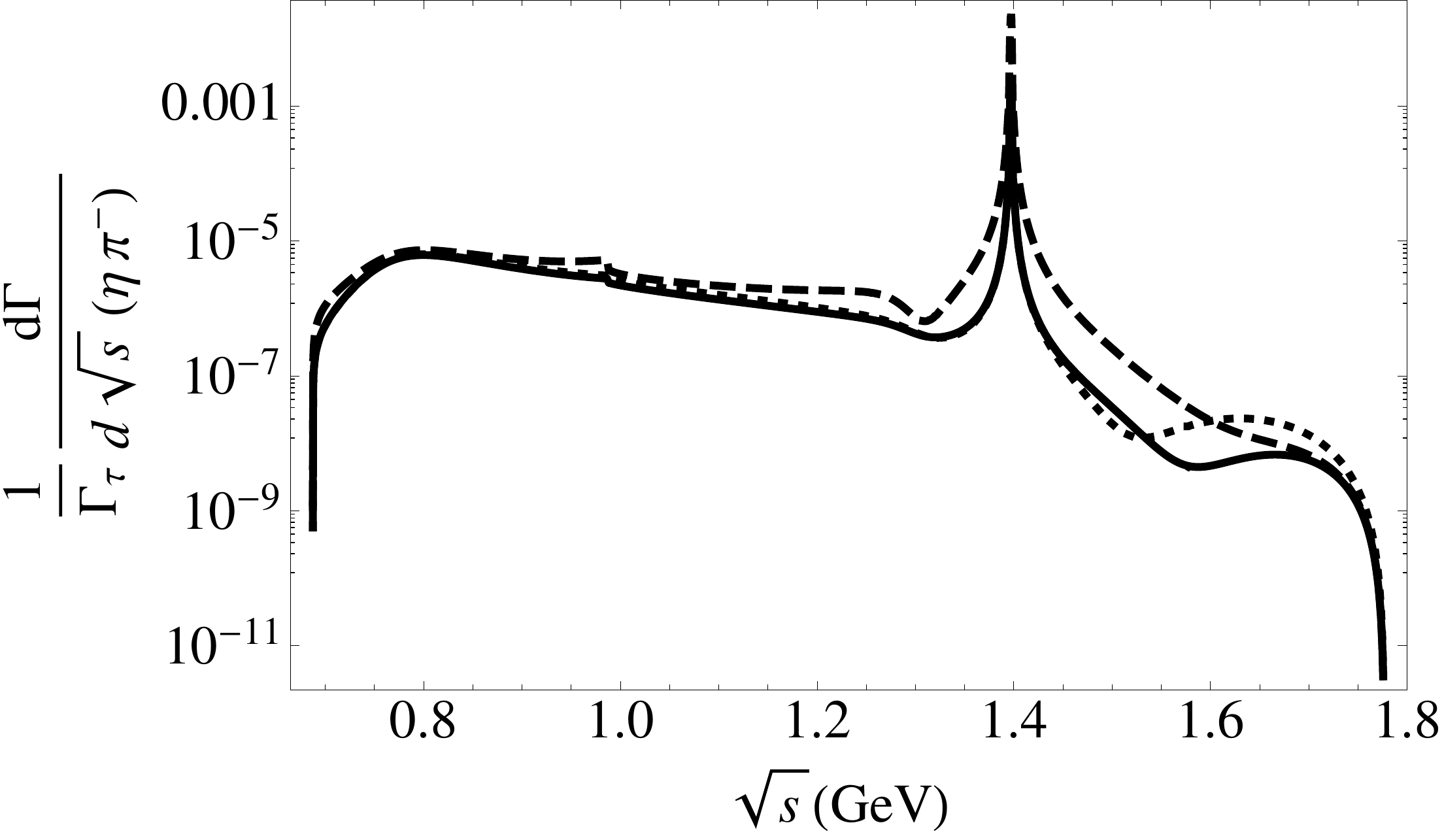}
    \includegraphics[scale=0.84]{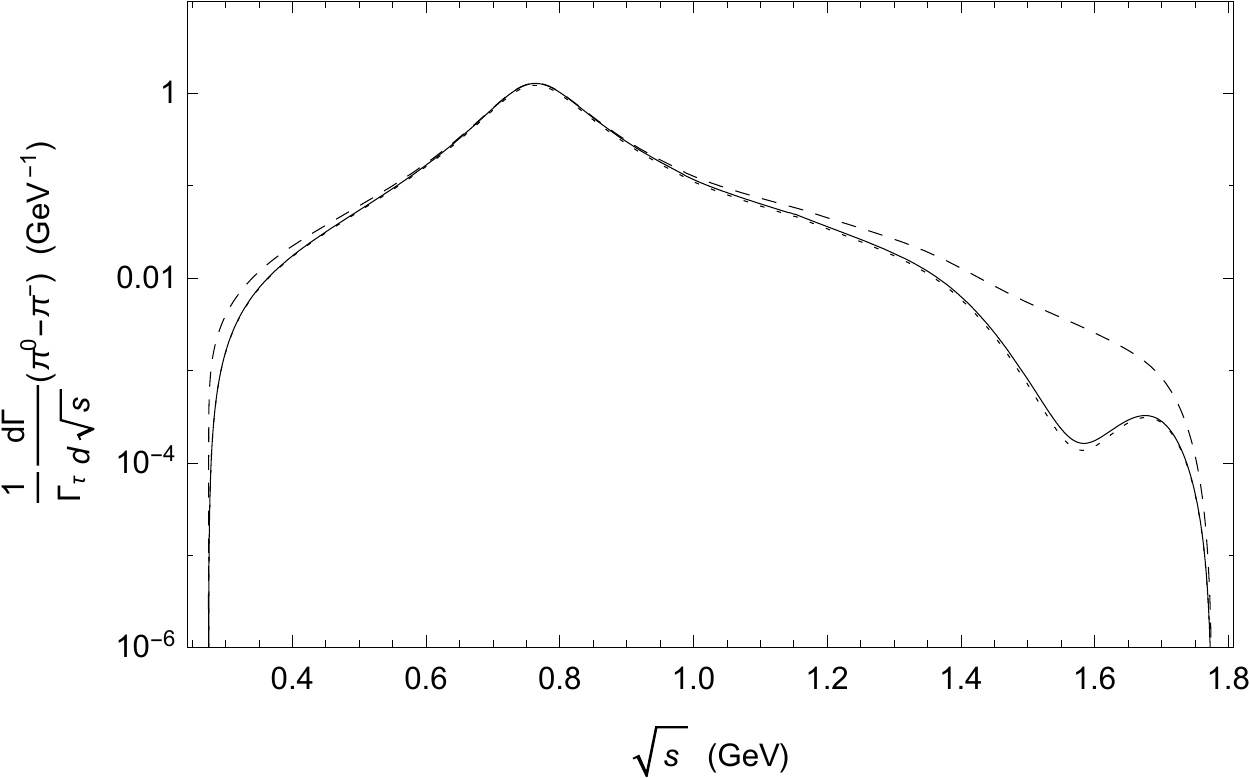}
    \caption{Normalized hadronic mass distributions (in GeV$^{-1}$ units; $\Gamma_{\tau}$ is the tau decay width). The upper plot corresponds to $\tau^- \to \pi^-\eta\nu_{\tau}$:  the solid line represents the SM contribution using \cite{Escribano:2016ntp} as the reference for form factors; the dotted (dashed) line corresponds to $\widehat{\epsilon}_S=0$ and $\widehat{\epsilon}_T=0.6$ (respectively, $\widehat{\epsilon}_S=0.004$ and $\widehat{\epsilon}_T=0$). The lower plot corresponds to $\tau^- \to \pi^-\pi^0\nu_{\tau}$ using the form factor model of Ref. \cite{Dumm:2013zh}: the solid line denotes the SM contribution; the dotted (dashed) line corresponds to $\widehat{\epsilon}_S=0$ and $\widehat{\epsilon}_T=-0.014$ (respectively, $\widehat{\epsilon}_S=1.31$ and $\widehat{\epsilon}_T=0$).  }  
     \label{hadr-pieta}
         \end{center} 
\end{figure*}

    The above expressions clearly illustrate the sensitivities of the two process under consideration to the effects of NS interactions.  In the case of the $\pi^-\pi^0$ final state the coefficient of the effective scalar form factor $\widetilde{F}^{P^0\pi^-}_S(s)$ is very suppressed and we could expect some sensitivity to the tensor interactions. For the $\pi^-\eta$ final state, the large mass difference between the hadrons in the final state, and the non-suppressed (by isospin breaking) scalar interactions,  enhances the sensitivity of the decay rates to that NS interaction.  Figure \ref{hadr-pieta} illustrates the sensitivity of the hadronic mass distributions to NS interactions: $(i)$ for the $\pi^-\eta$ ($\pi^-\pi^0$) channel, this observable is sensitive to small effects of scalar (tensor) interactions and $(ii)$ sizable distortions with respect to the SM contributions are observed in the region around the peak (respectively, close to the ends) of the spectrum in the case of $\pi^-\eta$ ($\pi^-\pi^0$) final states.

  \subsection{Decay rate}
  
    The observable in $\tau^- \to (\pi^-\eta, \pi^-\pi^0)\nu_{\tau}$ decays that is the most sensitive to the effects of non-standard interactions in our studies turns out to be the integrated rates. The partial rates $\Gamma$ can be obtained by integrating Eq. (\ref{hadmass}) over the whole kinematical range  (integration over a smaller kinematic range where the effects of NS interactions are larger can enhance the sensitivity). We define the following most sensitive observable:
    \begin{equation} \label{delta}
    \Delta \equiv \frac{\Gamma-\Gamma^0}{\Gamma^0} = \alpha \widehat{\epsilon}_S + \beta\widehat{\epsilon}_T+ \gamma  \widehat{\epsilon}_S^{\ 2}+\delta  \widehat{\epsilon}^{\ 2}_T ,
    \end{equation}
    where $\Gamma^0 = \Gamma( \widehat{\epsilon}_S= \widehat{\epsilon}_T=0)$ is the reference value computed using the form factors of Ref. \cite{Escribano:2016ntp} and can be identified with the SM decay rate.
 
   The coefficients of the quadratic form in Eq. (\ref{delta}) are displayed in Table \ref{coeff}: 
      \begin{table}[h!]
\centering
\begin{tabular}{|c|c|c|c|c|}
\hline
$P^-P^0$ channel & $\alpha$ & $\beta$ & $\gamma$& $\delta$ \\
\hline
$\pi^-\eta$ &$7.0 \times 10^2$ &$1.1 $ &$1.6\times 10^5$ &$21$ \\
$\pi^-\eta^{\prime}$ &$9.0\times 10^2$ &$-8.0\times 10^{-4}$ &$1.9\times 10^5$ &$0.1$ \\
$\pi^-\pi^0$ &$3.5\times 10^{-4}$ &$3.3^{+0.6}_{-0.4}$ &$2.2 \times 10^{-2}$ &$4.7^{+2.0}_{-1.0}$ 
 \\ \hline
\end{tabular}
\caption{\small Numerical values of coefficients in Eq. (\ref{delta}) \cite{Garces:2017jpz, Miranda:2018cpf} for different hadronic channels in $\tau^- \to \pi^- P^0\nu_{\tau}$ decays.  Error bars stemming from theoretical form factors is provided only for the $\pi^-\pi^0$ channel. }\label{coeff}
\end{table}
It is clear that decay rates of processes involving the $\eta$ ($\eta^\prime$) meson are sensitive to the effects of NS scalar interactions while the dominant and the most precisely measured semileptonic $\pi^-\pi^0$ channel is more sensitive to the effects of NS tensor interactions.
    
    \subsection{Constraints on NS interactions} 
    
\begin{figure*}[h!]
\begin{center}
    \includegraphics[scale=0.40]{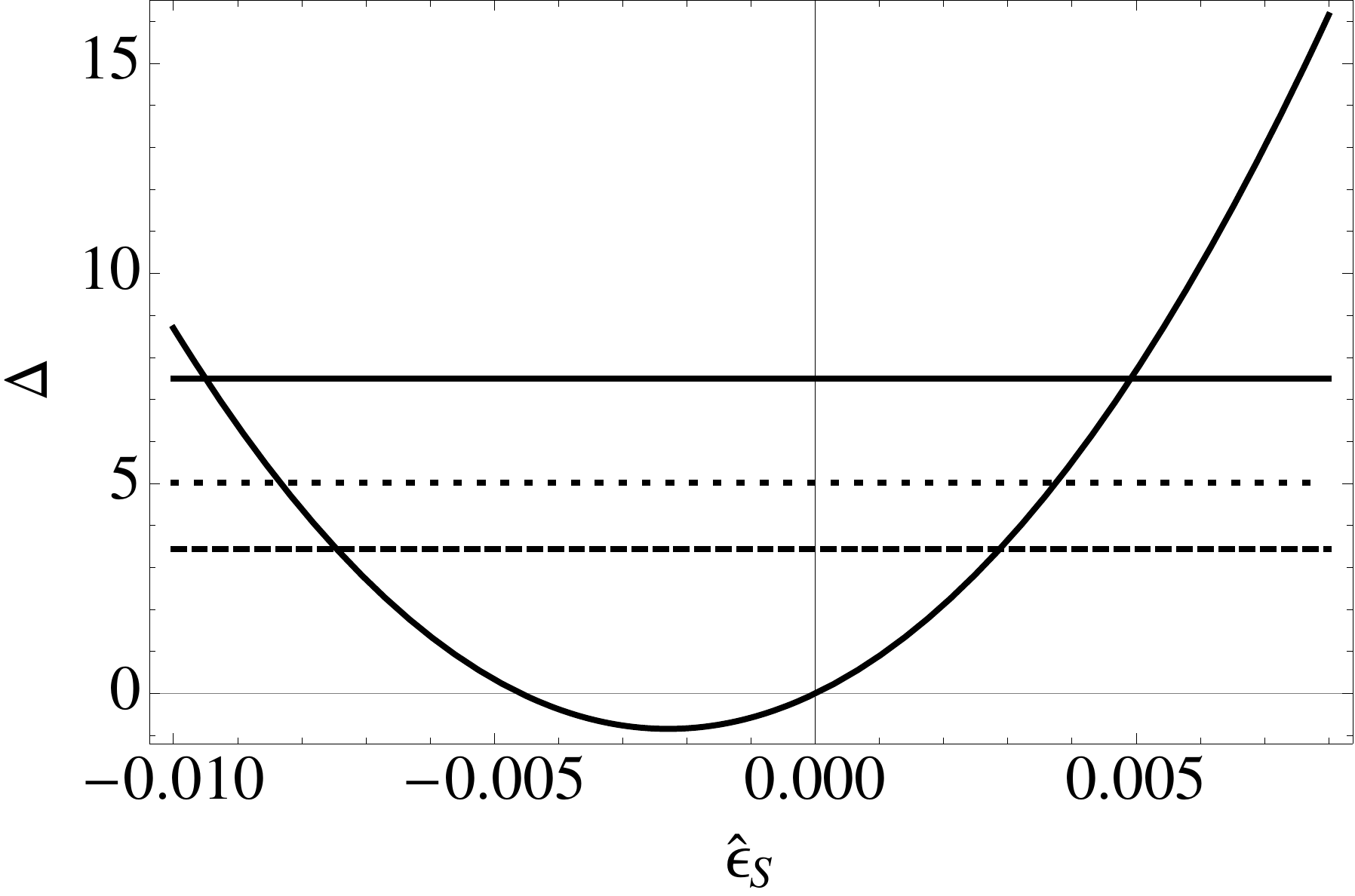}
    \includegraphics[scale=0.40]{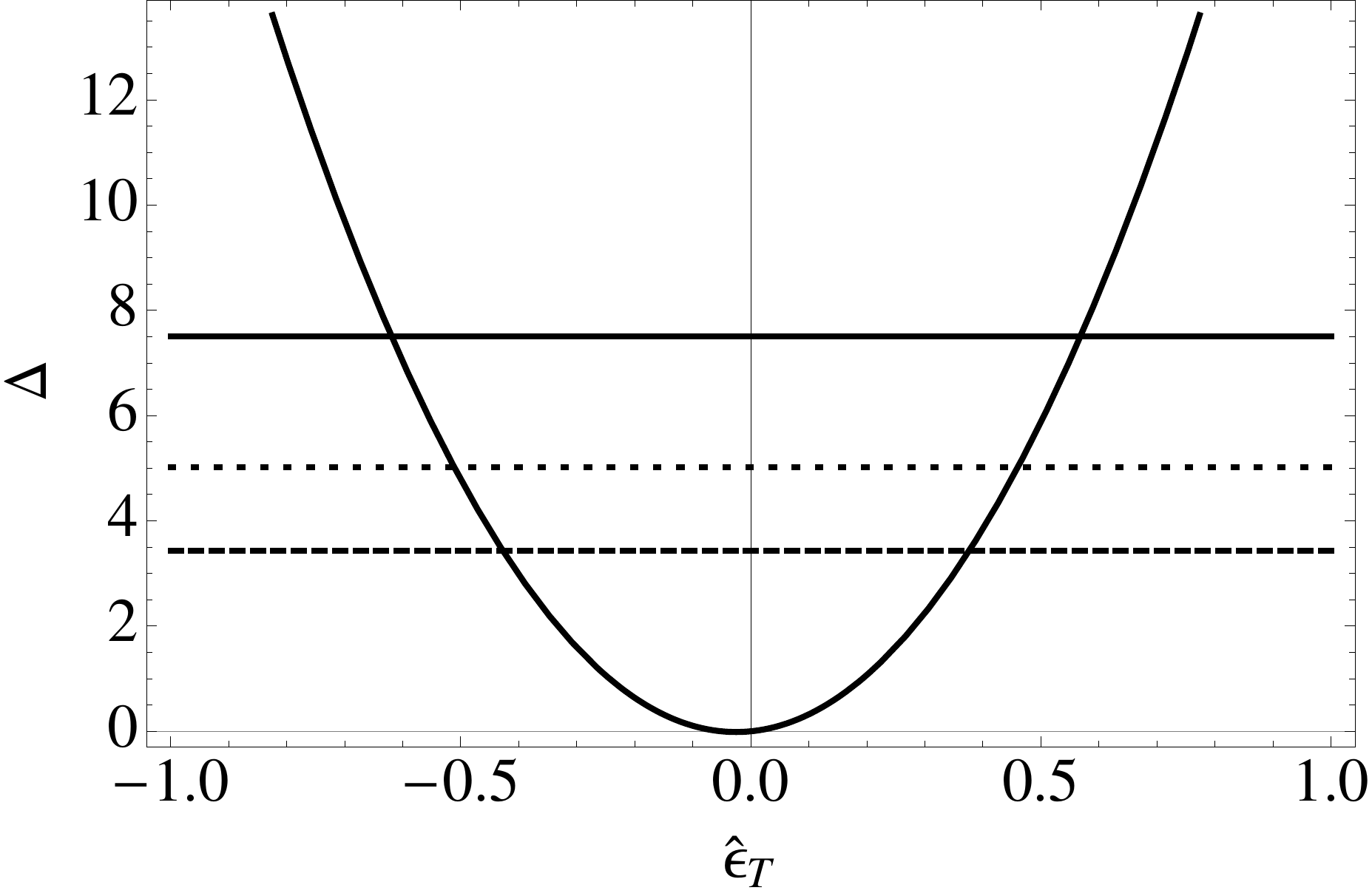}
     \includegraphics[scale=0.585]{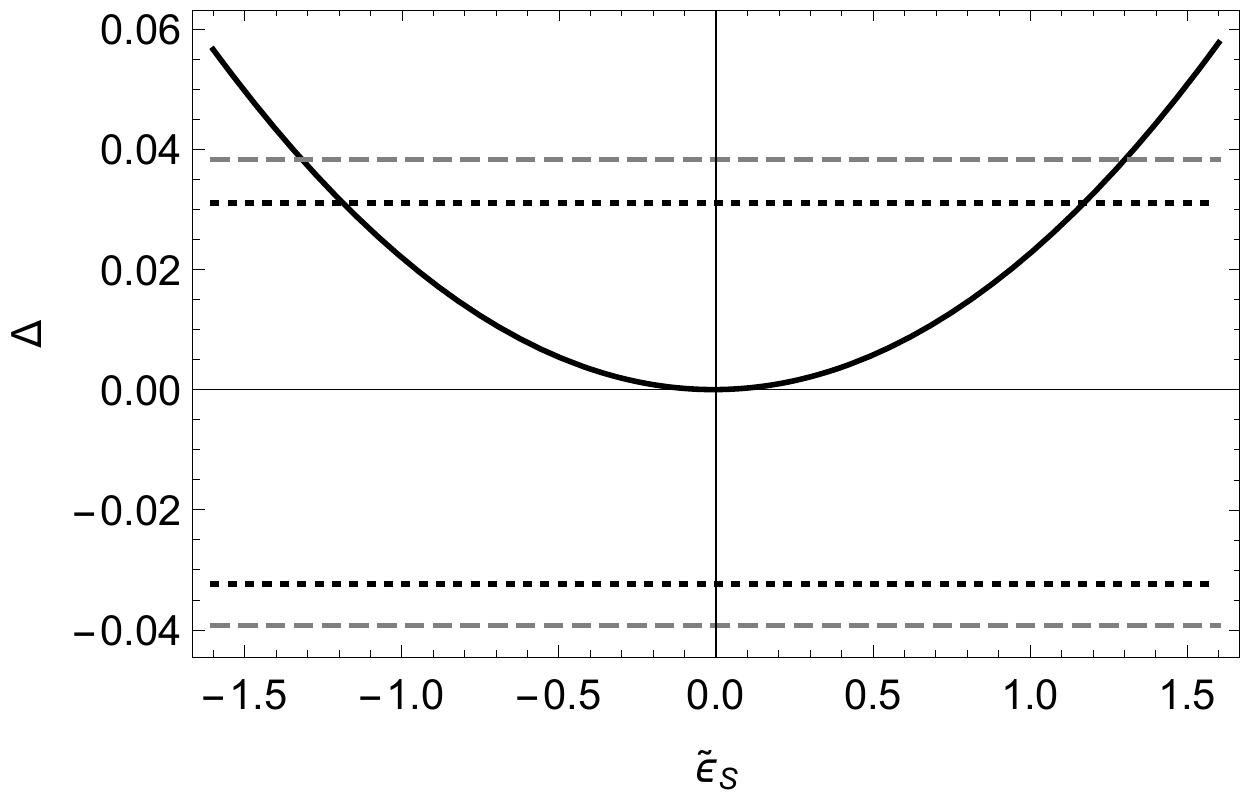}
    \includegraphics[scale=0.585]{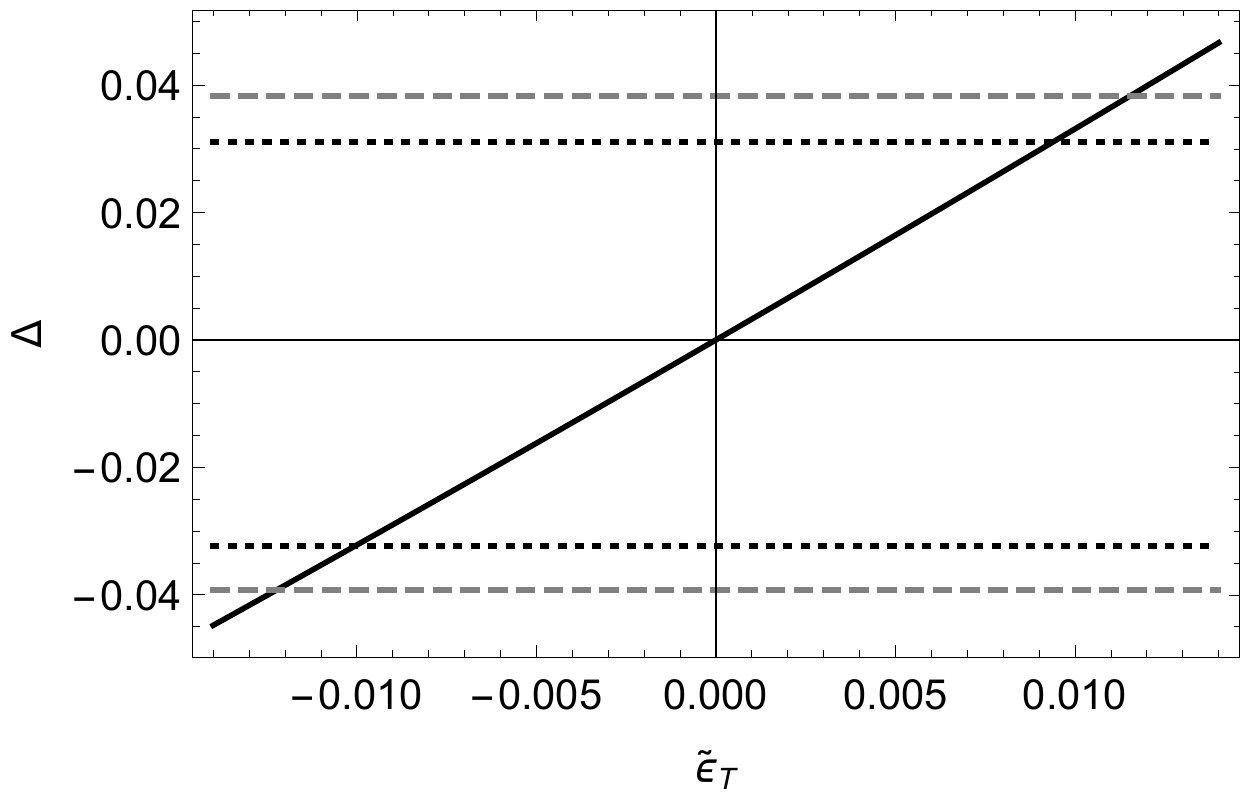}
    \caption{The parabola in the upper-half left-panel (right-panel) shows the constraints on $\widehat{\epsilon}_S$ when $\widehat{\epsilon}_T=0$ obtained from $\tau^-\to \pi^-\eta\nu_{\tau}$ decay (respectively, the constraints \cite{Garces:2017jpz} on $\widehat{\epsilon}_T$ when $\widehat{\epsilon}_S=0$);  the solid, dotted and dashed horizontal lines corresponds espectively to the upper limits obtained from CLEO \cite{Bartelt:1996iv}, BABAR \cite{Aubert:2008nj} and BELLE \cite{Hayasaka:2009zz} experiments. The solid lines in the lower-half represent the dependence on NS scalar (left) and tensor (right) couplings obtained from $\tau^- \to \pi^-\pi^0\nu$  decays according to Eq. (\ref{delta}); the band within dashed-lines corresponds to the region allowed  by the current experimental branching ratio\cite{Fujikawa:2008ma} within three-standard deviations; the hypothetical case that the error bars is reduced by a factor three (at Belle II) is shown by the band within dotted lines \cite{Miranda:2018cpf}. }  
     \label{eps-ST0}
         \end{center} 
\end{figure*}

    Figure \ref{eps-ST0}  exhibits the dependence of $\Delta$ on a single NS coupling (the other one is set to zero). This is compared (upper-half of the plot) to the upper limits on $\Delta(\tau^-\to \pi^-\eta\nu_{\tau})$ obtained from different $B$-factory experiments \cite{Bartelt:1996iv, Aubert:2008nj, Hayasaka:2009zz} which are represented by horizontal lines as indicated in Figure \ref{eps-ST0}. In the lower-half of the plot, a comparison is shown with the three standard-deviations error band of the measured branching fraction of $\tau^- \to \pi^-\pi^0\nu_{\tau}$ \cite{Fujikawa:2008ma} and with an hypothetical future improved measurement at Belle II \cite{Kou:2018nap}. In addition to improved measurements of branching fractions, better inputs for scalar and tensor form factors is required in order to further constraint (or discover!) NS couplings.
    
   \begin{figure*}[h!]
\begin{center}
    \includegraphics[scale=0.33]{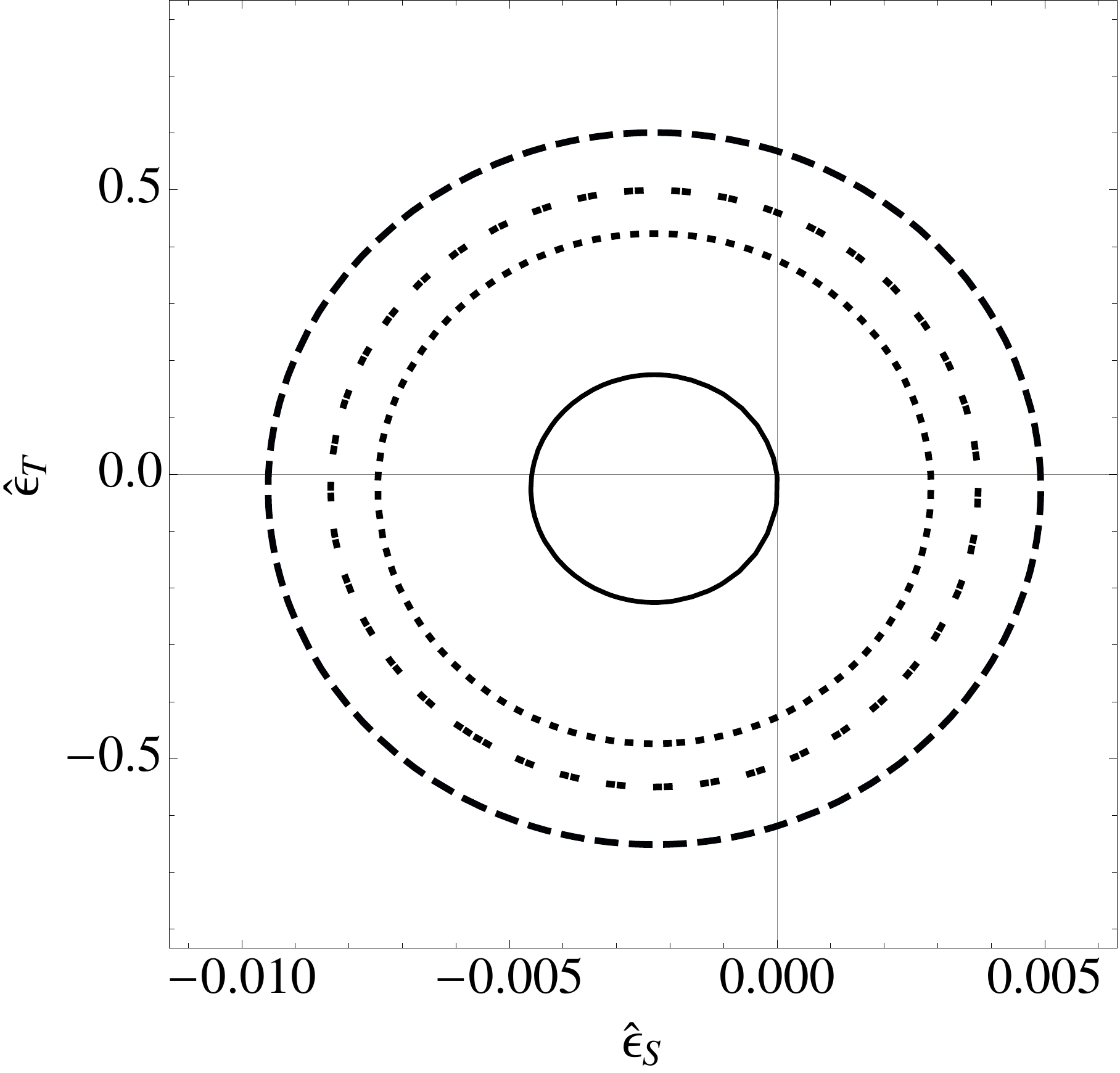}
    \includegraphics[scale=0.66]{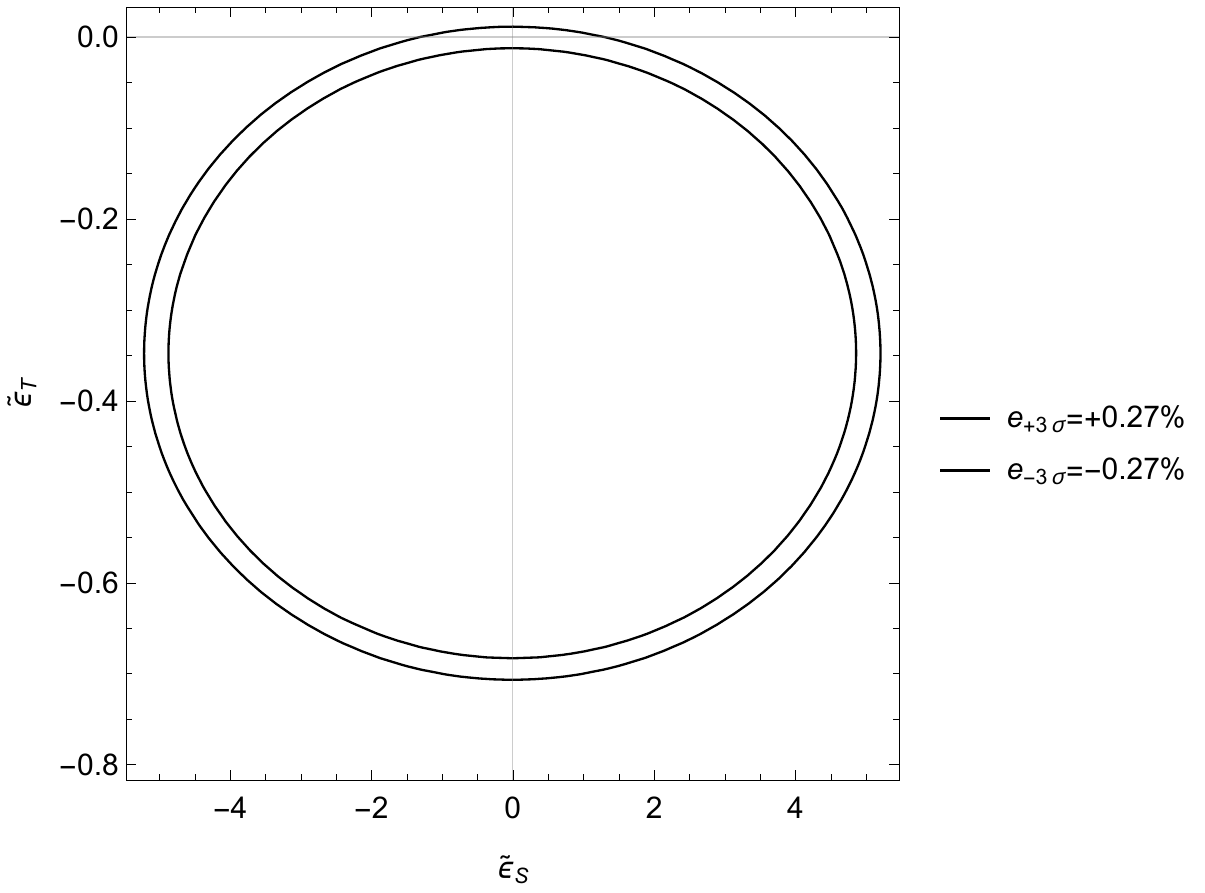}
    \caption{ Left: The allowed region  for $(\widehat{\epsilon}_S,\widehat{\epsilon}_T)$ \cite{Garces:2017jpz} from current upper limits on $B(\tau^-\to \pi^-\eta\nu_{\tau})$ by CLEO \cite{Bartelt:1996iv} (dashed line), BABAR \cite{Aubert:2008nj} (doubly-dotted line)and BELLE \cite{Hayasaka:2009zz} (dotted line) experiments; the solid line corresponds to $\Delta=0$. Right: allowed region for $(\widehat{\epsilon}_S,\widehat{\epsilon}_T)$ from measurements of  $B(\tau^- \to \pi^-\pi^0\nu_{\tau})$ \cite{Fujikawa:2008ma} including theoretical and experimental uncertainties at $3\sigma$  \cite{Miranda:2018cpf}. }  
     \label{epsS-epsT}
         \end{center} 
\end{figure*} 

Figure \ref{epsS-epsT} displays the constraints on both NS couplings derived from Eq. (\ref{delta}). The regions of parameter space for NS interactions are similar to the one in Figure \ref{eps-ST0}. Although only an upper limit is available for tau decays into $\pi^-\eta$ channel, combining the information on the allowed and suppressed semileptonic decays leads to very strong constraints on NS interactions, favouring couplings of $O(10^{-2}-10^{-3})$ for scalar and tensor couplings. The constraints that can be obtained from available measurements on the branching fractions without imposing an external input on one of the NS couplings are shown as intervals in Table \ref{results}. On the other hand, if we use the constraint $|\widehat{\epsilon}_S| \leq 8\times 10^{-3}$ from \cite{Cirigliano:2009wk}, we obtain \cite{Miranda:2018cpf}
\begin{equation}
\widehat{\epsilon}_T = \left(-1.3^{+1.5}_{-2.2}\right) \times 10^{-3}
\end{equation}
from a fit to the $\pi\pi$ spectrum of $\tau^- \to \pi^-\pi^0\nu_{\nu}$ decay measured by Belle \cite{Fujikawa:2008ma}, which is competitive to results obtained from other low-energy measurements \cite{Gonzalez-Alonso:2018omy} (although they are obtained from processes involving different leptonic currents).

  \begin{table}[h!]
\centering
\begin{tabular}{|l|c|c|l|}
\hline
channel         & $\widehat{\epsilon}_S$  & $\widehat{\epsilon}_T$& source \\ \hline
$\tau^-\to \pi^-\pi^0\nu_{\tau}$ & $(-5.2, 5.2)$ & $(-0.79, 0.013)$ & Ref \cite{Miranda:2018cpf} from BR \\ 
& $< 8\times 10^{-3}$ & $(-1.3^{+1.5}_{-2.2})\times 10^{-3}$ & Ref \cite{Miranda:2018cpf} from had. spectrum \\ \hline
$\tau^-\to \pi^-\eta\nu_{\tau}$ & $(-8.3, 3.7)$ & $(-0.55, 0.50)$ & Ref. \cite{Garces:2017jpz} from BR\\ 
& $(-6\pm 15)\times 10^{-3}$ & - & Ref. \cite{Cirigliano:2018dyk}  from BR \\ \hline
semileptonic $\tau$ & - & $(0.4\pm 4.6)\times 10^{-3}$ &Ref. \cite{Cirigliano:2018dyk} from  discrepancy of \\ 
&&& $e^+e^-$ vs $\tau$ $\pi\pi$ spectral functions \\ \hline
Other low-energy & $(1.4\pm 2.0)\times 10^{-3}$ & $(-0.7\pm 1.2)\times 10^{-3}$ & Ref. \cite{Gonzalez-Alonso:2018omy} \\
measurements &&& \\ \hline
\end{tabular}
\caption{ Constraints on NS scalar and tensor couplings obtained from semileptonic tau decays compared to the better constraints available from other beta decays. }\label{results}
\end{table}

These results show that NS scalar couplings are strongly constrained from the upper limits on $B(\tau^-\to \pi^-\eta\nu_{\tau})$, while NS tensor couplings are better constrained from the precision measurement of $B(\tau^-\to \pi^-\pi^0\nu_{\tau})$. They are similar to constraints obtained from other low energy observables as shown in Table \ref{results}, keeping in mind that different leptons are involved.

\section{Conclusions}
  Non-standard interactions arising from heavier degrees of freedom can manifest at low energies in semileptonic $\tau$ lepton decays as new four-fermion interactions characterized by a set of five ($\widehat{\epsilon}_{L,R,S,P,T}$) Wilson coefficients \cite{Cirigliano:2009wk}. {\it A priori} these new couplings do not obey lepton universality, thus semileptonic $\tau$ lepton decays offer a unique framework to constrain these NS couplings in view of current data and expected future improvements at $\tau$ lepton factories \cite{Kou:2018nap}. 
  
    In this talk we have presented a summary of our recent results on this subject \cite{Garces:2017jpz,Miranda:2018cpf} (a different approach using semileptonic $\tau$ lepton decays that reach similar conclusions is presented in \cite{Cirigliano:2018dyk}). We have shown that current upper limits on the branching fraction of the rare $\tau^- \to \pi^-\eta\nu_{\tau}$ decay, allow powerful constraints on  the scalar (second class) currents due to the suppression (by approximate isospin symmetry) of the Standard Model contribution \cite{Garces:2017jpz}. If this decay is measured at Belle II for the first time (see Ref. \cite{michel-tau2018}), it will allow to improve the present constraint reported here or (hopefully) observe the effects of NS scalar coupling. On the other hand, the very precise measurements of the hadronic spectrum and the branching fraction of the $\tau^- \to \pi^-\pi^0\nu_{\tau}$ decay, allow to set strong constraints on the NS tensor coupling. This is possible owing to the very strong suppression of the scalar form factor contribution in this case. 
   
   The expected improvements in the measurements/searches of these two semileptonic decays of the tau lepton at Belle II \cite{Kou:2018nap}, will allow to set some of the stronger constraints on NS scalar and tensor interactions. Of course, this will requiere an improvement in the description of the scalar  (for the $\pi^-\eta$ ) and the tensor (for $\pi^-\pi^0$) weak form factors. Hopefully, these expected  improvements will allow to constraint the  new weakly-coupled degrees of freedom to be heavier than the 10 TeV mass scale.

\section*{Acknowledgements}
  This work has been supported by Conacyt project CB-236394. I would like to thank Pablo Roig for several discussions, a fruitful collaboration and comments on this paper.




\nolinenumbers

\end{document}